\documentclass[sigconf]{acmart}

\AtBeginDocument{%
  \providecommand\BibTeX{{%
    \normalfont B\kern-0.5em{\scshape i\kern-0.25em b}\kern-0.8em\TeX}}}

\setcopyright{acmlicensed}
\copyrightyear{2024}
\acmYear{2024}
\acmDOI{XXXXXXX.XXXXXXX}

\acmConference[SE2030]{2030 Software Engineering
Workshop}{July 15--16, 2024}{Porto de Galinhas, Brazil}
\acmISBN{978-1-4503-XXXX-X/18/06}

\begin{document}

\title{A Road Less Travelled and Beyond: Towards a Roadmap for Integrating Sustainability into Computing Education}

\author{Ana Moreira}
\affiliation{NOVA LINCS \& NOVA University Lisbon \country{Portugal}}

\author{Ola Leifler}
\affiliation{Linköping University \country{Sweden}}

\author{Stefanie Betz}
\affiliation{
\institution{Furtwangen University, Germany}
\email{besi@hs-furtwangen.de}
}

\author{Ian Brooks}
\affiliation{
\institution{University of the West of England Bristol, U.K.}
\country{}
\email{Ian.Brooks@uwe.ac.uk}
}

\author{Rafael Capilla}
\affiliation{
\institution{Rey Juan Carlos University, Spain}
\email{rafael.capilla@urjc.es}
}

\author{Vlad Constantin Coroamă}
\affiliation{
\institution{Roegen Centre for Sustainability, Switzerland}
\country{}
\email{}
}

\author{Leticia Duboc}
\affiliation{
\institution{La Salle, Universitat Ramon Llull, Spain}
\country{}
\email{}
}

\author{João Paulo Fernandes}
\affiliation{
\institution{University of Porto, Portugal}
\country{}
\email{jpaulo@fe.up.pt}
}

\author{Rogardt Heldal}
\affiliation{
\institution{Western Norway University of Applied Sciences, Norway}
\country{}
\email{rogardt.heldal@hvl.no}
}

\author{Patricia Lago}
\affiliation{
\institution{Vrije Universiteit Amsterdam, Netherlands}
\country{}
\email{p.lago@vu.nl}
}

\author{Ngoc-Thanh Nguyen}
\affiliation{
\institution{Western Norway University of Applied Sciences, Norway}
\email{Ngoc.Thanh.Nguyen@hvl.no}
}

\author{Shola Oyedeji}
\affiliation{
\institution{Lappeenranta-Lahti University of Technology, Finland}
\email{}
}

\author{Birgit Penzenstadler}
\affiliation{
\institution{Chalmers University of Technology, Sweden}
\email{birgitp@chalmers.se}
}

\author{Anne Kathrin Peters}
\affiliation{
\institution{KTH Royal Institute of Technology, Sweden}
\email{akpeters@kth.se}
}

\author{Jari Porras}
\affiliation{
\institution{Lappeenranta-Lahti University of Technology, Finland}
\email{jari.porras@lut.fi}
}

\author{Colin C. Venters}
\affiliation{
\institution{University of Huddersfield, U.K.}
\email{c.venters@hud.ac.uk}
}

\renewcommand{\shortauthors}{Moreira et al. (2024)}

\begin{abstract}

Education for sustainable development has evolved to include more constructive approaches and a better understanding of what is needed to align education with the cultural, societal, and pedagogical changes required to avoid the risks posed by an unsustainable society. This evolution aims to lead us toward viable, equitable, and sustainable futures. However, computing education, including software engineering, is not fully aligned with the current understanding of what is needed for transformational learning in light of our current challenges. This is partly because computing is primarily seen as a technical field, focused on industry needs. Until recently, sustainability was not a high priority for most businesses, including the digital sector, nor was it a prominent focus for higher education institutions and society.

Given these challenges, we aim to propose a research roadmap to integrate sustainability principles and essential skills into the crowded computing curriculum, nurturing future software engineering professionals with a sustainability mindset. We conducted two extensive studies: a systematic review of academic literature on sustainability in computing education and a survey of industry professionals on their interest in sustainability and desired skills for graduates. Using insights from these studies, we identified key topics for teaching sustainability, including core sustainability principles, values and ethics, systems thinking, impact measurement, soft skills, business value, legal standards, and advocacy. Based on these findings, we will develop recommendations for future computing education programs that emphasise sustainability.

\textit{$[$The paper is accepted at the 2030 Software Engineering workshop, which is co-located with the FSE'24 conference.$]$}

\end{abstract}

%\begin{CCSXML}
%<ccs2012>
%  <concept_id>00000000.0000000.0000000</concept_id>
%  <concept_desc>Do Not Use This Code, Generate the Correct Terms for Your Paper</concept_desc>
%  <concept_significance>500</concept_significance>
  %<concept_id>00000000.00000000.00000000</concept_id>
%  <concept_desc>Do Not Use This Code, Generate the Correct Terms for Your Paper</concept_desc>
%  <concept_significance>300</concept_significance>
  %<concept_id>00000000.00000000.00000000</concept_id>
 % <concept_desc>Do Not Use This Code, Generate the Correct Terms for Your Paper</concept_desc>
 % <concept_significance>100</concept_significance>
  %<concept_id>00000000.00000000.00000000</concept_id>
%  <concept_desc>Do Not Use This Code, Generate the Correct Terms for Your Paper</concept_desc>
%  <concept_significance>100</concept_significance>
%</ccs2012>
%\end{CCSXML}

%\ccsdesc[300]{Do Not Use This Code~Generate the Correct Terms for Your Paper}
%\ccsdesc{Do Not Use This Code~Generate the Correct Terms for Your Paper}
%\ccsdesc[100]{Do Not Use This Code~Generate the Correct Terms for Your Paper}


\maketitle

\section{Introduction} \label{introduction}
ICT systems and services are the backbone of modern society, deeply embedded in every aspect of our lives, from health and commerce to communication, education, energy, finance, defence and beyond. While these complex, large-scale, software-intensive systems might be vital enablers for sustainable development, their development and utilisation also present significant sustainability challenges. Indeed, the impact of ICT on economic, societal, and environmental sustainability is extensive, influencing areas such as social equity, carbon emissions, and resource production and consumption~\cite{Creutzig:2022aa}, and thus is particularly challenging to assess~\cite{bremer23:assessing}.

The complexity of our world has grown significantly, leading to an increased reliance on computers to help us understand it. For example, considering that 80\% of the ocean remains uncharted territory, we have nevertheless started deep-sea mining, fish farms and offshore wind farms. Amidst such ventures, where the potential impacts are largely unknown, data collection, analysis, and monitoring are imperative~\cite{IoTJournal}. However, the use of large-scale data collection and advanced AI tech is generally not directed towards such goals, as societal goals tend to be directed towards economic short-term advantages~\cite{Creutzig:2022aa}.

In short, our current deep and interlinked challenges call for ``rethinking growth, rethinking efficiency, rethinking the state, rethinking the commons, and rethinking justice''~\cite{McPhearson:2021aa}, which in turn would call for ``redirecting digitalisation toward stabilisation of human and planetary systems''~\cite{Creutzig:2022aa} and move away from ``narrow techno-economic mindsets and ideologies of control''~\cite{Stoddard:2021aa}.
This forward-looking paradigm must prioritise minimising environmental impact, championing social equity, and fostering economic resilience. It is, therefore, essential to equip professionals and students with the knowledge, skills, and tools needed to design, develop, and manage software and systems for such a paradigm. 

Education is a key leverage point in enabling transition mindsets and providing opportunities to learn about alternative pathways forward~\cite{UNESCO:2017aa}. As educators, we can play an important role in designing educational frameworks that instil a sustainability mindset among current and future ICT professionals. Hence, the challenge is: \textit{How can we infuse sustainability principles and cultivate essential skills and competencies in an already crowded computing curriculum to nurture the future generation of software engineering professionals with a sustainability mindset?}

During the past three years, we focused on laying the foundation for integrating sustainability into computing education, while identifying current professionals' training needs. This involved conducting two comprehensive studies: one examining how academia addresses sustainability in teaching \cite{peters2024sustainability}, and the other engaging with industry to understand their stated needs, challenges, and practices when handling sustainability in their businesses \cite{heldal2024sustainability}.
 
 Drawing from the findings of these studies, we now propose initial insights to inform the design of future programmes for integrating sustainability in computing education and highlight open questions for a future roadmap. 
 
Such a roadmap must differ from traditional approaches, emphasising broader societal challenges and preparing students to navigate complex, wicked problems by fostering personal development, self- and ethical awareness, and the capacity to respond to complex problems with care and empathy. 

This paper is structured as follows: Section~\ref{sec:background} offers an overview of the two studies that constitute the groundwork for the roadmap for teaching sustainability to ICT students and professionals. Section~\ref{sec:insights} provides a preliminary classification of the
topics that could be considered in sustainability training and education. Section~\ref{sec:takeaways} critically examines the proposed topics in light of the results of the literature review on sustainability in computing education, highlighting open questions within each topic area.
Finally, Section~\ref{sec:conclusions} concludes the paper and discusses some of the challenges to propose a sustainability education curriculum with the power to reform current computing education expectations.

\section{Background} \label{sec:background} 
The work presented in this paper is based on two prior studies, one studying how sustainability has been considered and implemented in education~\cite{peters2024sustainability} and the other studying the needs and challenges of industry~\cite{heldal2024sustainability}. We briefly outline the outcomes of these studies in the following subsections as a foundation for this paper.

\subsection{An overview of academic endeavours} \label{sec:slr}
Based on a systematic study of the scientific literature on sustainability in computing education work we explored the current state of practice and which changes would be called for given literature on education for sustainable development and transformative education~\cite{peters2024sustainability}. 

The literature review was based on an initial set of 473 papers from a database search covering the period of 2000--2022, reduced to 45 papers to specifically include papers addressing sustainability in computing education, and later extended through snowballing to 89 papers. This set was analysed with respect to the specific learning goals activities and research methods employed. The analysis made comparisons of learning goals and activities to what we found in the literature on learning for sustainable development, and compared research methods to methods established through ACM empirical standards.

The review concluded with recommendations and open questions around four themes: (1) the research on sustainability in computing education in light of the severity of our challenges and demands for unprecedented change, (2) connections to research on equality, justice, norms, values and power, (3) implications of current approaches to curriculum development and directions for future work, and (4) limitations of existing research and outlook.

Through these themes, the review assessed approaches to include sustainability in computing education and provided recommendations for change around several axes: to make more meaningful, substantial changes that can engage with the seriousness of the issues, to move away from a curricular alignment with existing business profiles and towards exploration of new skills and attitudes, to help students take action while being grounded in a mature relationship to difficult emotions, to connect explicitly with work on understanding power relations, equality and justice, to escape the ``narrow techno-economic mindset''~\cite{Stoddard:2021aa} represented by computational thinking, and to ensure that students can cope in a mature way with material that challenges them emotionally~\cite{Pihkala:2022aa}.

As the literature review on computing education was ambivalent regarding whether existing practices could offer specific guidance in light of this, it ended with a series of open questions.

\subsection{An industry perspective on the needed competencies and skills}  \label{sec:industry}
In previous work, we explored industry's views on the sustainability education and training of software engineers~\cite{heldal2024sustainability}. To do so, we conducted interviews and focus groups with IT and sustainability professionals from 28 organisations across various sectors and countries.  We aimed to learn about (i) their interest in sustainability; (ii) their sustainability goals; (iii) the sustainability-related competencies and skills needed to achieve these goals.

Concerning their \textbf{overall interest in sustainability}, we asked companies about four perspectives: business, customers, shareholders, and stakeholders. From the business point of view, they highlighted the business opportunities and increased competitiveness brought by sustainability. They also felt they had to respond to customers' environmental concerns and, to a lesser extent, social concerns. Regarding shareholders, the economic benefits of the companies were the unsurprising main interest, but social concerns were also mentioned. Finally, employees' interests, the media and the regulatory framework were also mentioned as drivers to address sustainability within their businesses.

When discussing their \textbf{sustainability goals}, interviewees sought to make their processes and products more sustainable, recognising the need for a culture change in their employees and customers. Achieving such goals required collaboration with partners and external entities, and the transformation of work processes, and technical tools.  However, they noted that employees lacked an understanding of sustainability and still saw it as a trade-off against profit. Interviewees also complained about economic barriers and inadequate policies, when trying to reach their sustainability goals. 

When it came to the \textbf{sustainability-related competencies and skills}, interviewed individuals acknowledged the importance of soft skills and technical competencies, but six companies faced difficulties in hiring the right talent with suitable sustainability-related skills. They wished their employees to have better communication skills and knowledge about sustainability metrics, so, they often resorted to training courses, sustainability consultants and collaborations with universities. 

Finally, noting how interviewees referred to the different sustainability dimensions was also interesting. Most of them talked about environmental issues, such as carbon emissions and climate change. Those discussing the economic dimension often meant the company's ability to maintain its profits over time. The social dimension was mostly addressed through initiatives for employees' well-being, customer satisfaction, and societal welfare. Finally, for the technical aspects, some mentioned well-known software qualities, such as reusability, robustness, and security. Surprisingly, only a minority of companies mentioned the UN Sustainable Development Goals (SDGs).

From the insights we got from these interviews, we highlight that professionals need: (1) a comprehensive understanding of the multiple dimensions of sustainability and their alignment to business opportunities; (2) proficiency in soft and sustainability skills to promote awareness among employees, customers, and collaborators; and (3) tools and metrics to integrate sustainability into their businesses' products and processes. The Code book \footnote{Codebook access link: \href{https://bit.ly/3wgt0dp}{https://bit.ly/3wgt0dp}} of the interviews is shared as supplementary material to help readers better understand our insights.

\section{Insights for a future education programme}\label{sec:insights}
An increasing number of Higher Education Institutions (HEI) recognise the need to align education with the need for societal transformations that our challenges demand. While it is acknowledged that the integration process is challenging~\cite{fiselier2018critical}, there is some guidance available for educators in computing, such as the European sustainability competence framework (GreenComp)~\cite{RePEc:ipt:iptwpa:jrc128040} and the UK QAA/HEA guidance on incorporating Education for Sustainable Development into the curricula~\cite{QualityAssuranceAgencyforHigherEducation2021}. 

Based on our interpretation of the results from our two studies~\cite{heldal2024sustainability,peters2024sustainability} we constructed a preliminary classification of topics to be taught when designing a curriculum for future courses in computing integrating sustainability. This classification reflects our perspective on key considerations, stemming from intensive discussions in regular workshops among the authors, alongside our extensive research and teaching experience in the field. It is worth noting that while this classification is not exhaustive, it underscores pivotal areas we believe merit focus. Our proposal goes beyond the core sustainability knowledge outlined in~\cite{glasser2016toward, RePEc:ipt:iptwpa:jrc128040} which proposes a list of the minimum competencies designed to facilitate transformative systems' changes to enhance the overall quality of life. Additionally, our focus extends to encompass soft skills, technical sustainability, the business rationale for sustainability, assessing sustainability impacts and metrics, ethical considerations and values, legal requirements and standards, and advocacy and lobbying efforts. In the following, we describe the importance of each topic, how it relates to our studies, and identify education opportunities. 

\subsubsection*{Core sustainability concepts} 
Building a sustainable future demands that communities and individuals are cognisant of sustainability principles and the ramifications of human actions on the interconnected environmental, economic, and social dimensions of sustainability. Given the fundamental role of ICT systems and services in modern society, sustainability must be considered an essential cornerstone for every IT product or service. 

We observed a lack of consensus among industry professionals regarding the definition of sustainability. Fourteen organisations (50\% of the total) cited a significant challenge in understanding sustainability concepts. Additionally, while 10 organisations appeared to view sustainability primarily through the lens of environmental concerns, a smaller number also mentioned technical, social and economic concerns.  Furthermore, 12 organisations explicitly acknowledged sustainability as intertwined with multiple dimensions. Surprisingly, only four mentioned the Sustainable Development Goals as pertinent to their operations.

Insufficient comprehension of sustainability's core concepts may lead to various challenges: perceiving it as an optional feature rather than a crucial aspect; struggling to collaborate with sustainability experts, and might not fully see the value of it; shying away from public debate on technology and sustainability, not advocating policy changes; seeing it as an auxiliary skill to IT rather than an integral component. 

Education plays a pivotal role in providing knowledge of sustainability's basic principles, concepts, and models. This includes defining and scoping sustainability, dispelling prevalent misconceptions and myths, presenting current statistics, elucidating key concepts (e.g., dimensions of sustainability~\cite{penzenstadler2014infusing}, sustainable development goals\footnote{\href{https://sdgs.un.org/goals}{https://sdgs.un.org/goals}}), and introducing various models for contextualising sustainability (e.g., the doughnut model~\cite{Raworth2017}, the nine planetary boundaries~\cite{Richardson2023}, and orders of impact~\cite{hilty2015ict}).

\subsubsection*{Values and Ethics} 
Ultimately, values and ethics are fundamental concerns to making the world a fair and equitable place~\cite{shearman1990meaning}. While ethics are culturally agreed-upon moral principles, ``values make no moral judgment''~\cite[p.~113]{8693084}. Currently, our society relies on software systems to communicate worldwide and operate utilities providing the basics for human life (e.g., complex medical machines, nuclear plants, and electrical grids). Such systems focus on functionality and quality attributes, such as usability, availability, and security, but they do not respect basic human values, such as social justice, transparency, or diversity~\cite{schwartz2007basic}. Sustainable systems, however, should be aligned with core human values.

Over 57\% of the interviewed organisations reveal that their customers and stakeholders want to protect the environment, and almost 30\% are interested in focusing on sustainability due to moral concerns and social matters, resulting in the need for sustainability-value alignment of their business. Therefore, it is important that IT and professionals are guided by a clear code of ethics, such as the ACM code of ethics~\cite{gotterbarn1997software} to produce socially and environmentally responsible systems. 

As educators, we call for ethics to be a standard part of software engineering~\cite{gotterbarn2017acm}. With regards to values, all three approaches of user-centred design, user-experience design, and values-sensitive design tackle more than typical software qualities, but they are still far from addressing core human values~\cite{Alidoosti2022}. Value-driven methods, known in HCI and information systems, can be used in business analysis and requirements engineering, but they offer little guidance for the later stages of development. Some emerging works take a human-values view (e.g.,~\cite{burnett2016gendermag} used to discover gender bias in software; or~\cite{Alidoosti2022b} incorporating ethical values in software design), but more is still required to address human values systematically. The good news is that software development methods could be adapted to handle human values. For example, the Karlskrona Manifesto on Sustainability Design~\cite{Karlskrona}, or participatory design techniques can be taught to ensure that end-user values are taken into account.

\subsubsection*{Systems thinking} 
Systems thinking is a term used to describe several knowledge traditions that seek to understand the underlying, general mechanisms of how the world works and how to understand the dynamics and principles necessary to grasp those systems~\cite{Ramage:2020aa}. Some contemporary forms of systems thinking traditions, such as cybernetics, general systems theory, chaos theory, and system dynamics invite thinking about social and ecological systems in terms of drivers and dynamics that are common to many systems but need to be understood as phenomena on their own terms. Based on those traditions, systems thinking becomes a new lens, a new orienting principle for understanding the world and pathways of change in it~\cite{Linner:2021aa}. 

Other forms of systems thinking such as Soft Systems methods and Critical System Heuristics rather call into question the frames of reasoning and rationale for delimiting, for example, problem scopes, relevance of knowledge claims, success metrics and justifications~\cite{Ulrich2010}. Based on those traditions, systems thinking becomes a tool for critically reviewing the requirements of systems and finding better boundaries when making judgements in reasoning about systems~\cite{Duboc:2020aa}.

Systems thinking can also be seen as a means of becoming able to care for the wider ecosphere and humanity as a part of it, and to make design decisions based on understanding the implications of such interconnectedness. Here, systems thinkers such as Bateson~\cite{Gregory:1979aa} have a great deal in common with indigenous scholars such as Kimmerer~\cite{Kimmerer:2013aa} who argues that thinking in terms of systems implies that we must be able to cultivate reverence for the primacy of life, and foster an inner development of people to see themselves as part of a greater system and wish to act in such a way as to care for the conditions for thriving of human beings within a regenerating web of life~\cite{Stalne:2022aa}. Based on such traditions, systems thinking becomes a means to broaden the view of which types of competencies are valuable to orient education towards and to explicitly include the inner development of your capacity to be, care for, and act in, the complex adaptive systems of life.

Several of the interviewed organisations recognised the importance of systems thinking, emphasising the need for a holistic view of factors and interactions that could contribute to a better possible outcome~\cite{heldal2024sustainability}. This awareness presents an opportunity for educators, to integrate systems thinking into computing education. By adopting this approach, educators can lay the groundwork for fostering more just and sustainable futures in computing, moving away from current paradigms that may hinder progress in addressing societal challenges~\cite{Becker:2023aa}. In doing so, computing needs to be re-situated as a practice that seeks to redress problems in terms of those who are primarily affected negatively by contemporary exploitative and destructive practices~\cite{Becker:2023aa}.

\subsubsection*{Sustainability Impacts and Measurements} 
The role of metrics and other sustainability indicators are key to evaluating to what extent an organisation or a particular sustainability initiative achieve certain SDGs. Companies are adopting sustainable practices that have to be measured with adequate indicators aimed to prove they meet the relevant SDGs. While certain domains (e.g., energy or transportation) have well-established indicators aimed at evaluating the sustainability of a solution, others do not. Consequently, many organisations have difficulties assessing sustainability and they need to define appropriate indicators. Existing approaches to assess sustainability include techniques and tools such as the SDG impact assessment tool~\cite{chalmers2019sdg}, the SAF sustainability assessment toolkit \cite{SAFToolkit,Lago2019}, and frameworks such as SuSAF~\cite{duboc2020requirements}. In addition, public and private universities are being increasingly ranked according to sustainability indicators (e.g. Times Higher Education Impact  Rankings), based on a set of predefined metrics\footnote{\href{https://www.timeshighereducation.com/impactrankings}{https://www.timeshighereducation.com/impactrankings}}.

From our analysis of the companies, we found six organisations highlighted the lack of metrics to understand the direct impacts of the product/system adopting sustainable solutions, while another company observed there is a lack of awareness on whether the solutions adopted are sustainable enough, partly because of lack of metrics. These and other technical challenges aimed at calculating the carbon or energy footprint of sustainable solutions are one of the reasons to demand specific training on well-defined KPIs that can justify their achievement of a certain level of sustainability. Nevertheless, many companies today (e.g. Google, Microsoft, CGI) rely on public sustainability reports where they deliver a set of sustainability indicators achieved by their products. However, while many of these indicators concern energy savings (energy indicators being popular nowadays), it is unclear to what extent the metrics used by each company are standardised. Therefore, organisations providing sustainability indicators need to make public how these metrics and other sustainability indicators have been computed. For instance, Google offers a carbon footprint calculator\footnote{\href{https://cloud.google.com/carbon-footprint?hl=en}{https://cloud.google.com/carbon-footprint?hl=en}}. 
Nevertheless, it is important to note that monitoring the sustainability impact of software systems requires a suitable dataset with sufficient quality. How to collect the right data and collect the data sustainably remains to be a challenge to be addressed. For example, while collecting marine data is expensive~\cite{IoTJournal}, many organisations have repeatedly collected the data already obtained by others due to the fragmented data sharing systems within the domain~\cite{lima2022marine}. The environmental cost of operating computing platforms to handle the data is an important metric to monitor~\cite{kumar2012green}.

Consequently, the use and adoption of existing sustainability metrics in various domains (e.g. climate change measures) and sustainability dimensions (e.g. technical and environmental) brings new opportunities for students to be taught on those metrics and rankings and these topics should be included in CS studies. The sustainability assessment toolkit (SAF)~\cite{SAFToolkit} is one of the examples that can be used to evaluate sustainability-related requirements. Toolkits like this can be used to train CS students in the evaluation of sustainability concerns. 

\subsubsection*{Technical sustainability} 
While there is no consensus within the field of software engineering as to how sustainability should be defined or understood, technical sustainability can generally be defined as the capacity of the software system to endure in changing environments~\cite{becker2015sustainability,VENTERS2018}. Emerging views have evolved to argue that technically sustainable software is that which is "\textit{explicitly designed for continuous maintainability and evolvability without incurring prohibitive technical debt and a negative
impact on the dimensions of sustainability}"~\cite{VENTERS2023}. 
The results from our analysis of industry needs, confirm that for the technical dimension, the interviewees viewed sustainability in relation to quality attributes of IT products and services, such as reusability, robustness, security, etc~\cite{heldal2024sustainability}. This strongly aligns with previous studies investigating how software engineering professionals understand sustainability, highlighting traditional software quality concerns such as maintainability and extensibility~\cite{chitchyan2016,Groher2017}. 

From our analysis, we identified several fundamental software engineering skills and competencies that are missing from companies including the application of software architecture in the design of sustainable software systems, and an understanding of and the application of software metrics to evaluate technical sustainability in a range of diverse application domains~\cite{heldal2024sustainability}. It has long been argued that software architectures are fundamental to the development of technically sustainable, i.e., long-living, software systems, as they are the primary carrier of architecturally significant requirements (ASRs) and influence how developers are able to understand, analyse, test, and evolve a software systems~\cite{Lilienthal2019}. 

Software metrics provide a quantifiable measurement of software characteristics. While there is a large number of metrics to track software development to evaluate software quality and to certify software products, there is still a lack of information and understanding on how to choose the most suitable metrics in a particular context to evaluate technical sustainability~\cite{Seacord2003}. Understanding which software metrics contribute to software quality assessment and how to automatically integrate them into a software development pipeline is critical. For example, Org. 18 mentions they use some metrics, but they lack automatic processes to evaluate the quality of their software systems. Understanding how to design software architectures and apply software metrics can help to produce more software systems that are technically sustainable~\cite{VENTERS2023}.

Despite a range of software engineering methodologies and methods --- waterfall, iterative waterfall, spiral, v-model, rational unified process, rapid application development, test-driven development, object orientation, DevOps, extreme etc. --- that have emerged over the last fifty years, it is argued that the field of software engineering has still not fully understood how to successfully construct successful software systems on a repeatable basis that does not result in software projects being significantly over time and budget or failing as a result of their size and complexity~\cite{Jacobson2018, Gharbi2019}. As a result, what remains is the need to align these perspectives with a core set of software engineering skills and competencies that enable quantifiable improvements in code and architectural design, and software comprehension, and maximise its reuse, leading to standardised software development practices and sustainable software systems. There is still a weak connection that proves that many of the existing metrics can be used to quantify technical sustainability, and this gap must be included in CS courses to guide students about selecting the most suitable metrics or a combination of them that can be chosen to measure sustainability concerns.

\subsubsection*{Soft skills} 
Soft skills are a term used to describe various generic skills or competencies that could be used in a variety of job contexts as opposed to hard skills that are linked with the technical expertise needed for the work \cite{olmstead1974research}. Marin-Zapata et al.~\cite{Marin-ZapataSaraIsabel2022Ssdw} studied the conceptualisation of soft skills in the literature and suggests, based on approaches found from the literature, that soft skills have two main components: intrapersonal and interpersonal. While the soft skills research initially focused on interpersonal skills, i.e., people and social skills, like communication and leadership, the recent research emphasises also the ability to manage oneself, i.e., intrapersonal skills.  Marin-Zapata et al.~\cite{Marin-ZapataSaraIsabel2022Ssdw} present examples of soft skills focusing on these two main components but does not provide a comprehensive categorisation of various soft skills.

From our analysis of company needs, we identified several skills that either exist in or are missing from a company. We divided these skills into profession-specific hard skills, soft skills, and sustainability-specific skills. Categorising the findings of the soft skills as interpersonal and intrapersonal skills reveals that companies both have and need more intrapersonal skills. Personal qualities like common sense, critical reflection, and problem-solving skills, as well as values and ethics, confidence and ways of working, and stress management, are valued by the companies. The identified interpersonal skills include Collaboration, Communication, Leadership, and Influencing skills, amongst others.

Although the importance of soft skills is increasingly recognised in computing degrees, educators often struggle to fit the topic properly into their classes. This difficulty may arise for several reasons. Sometimes, computing teachers do not have the right knowledge and tools to teach soft skills to students effectively. On other occasions, they may feel that the time they have to cover the technical curricula is already too short, resulting in a weak integration of soft skills into their teaching (e.g. without giving the students the required time to reflect on and practice their soft skills). Designers of computing programs should collaborate with other courses where soft skills play a more critical role, such as management, leadership, and social psychology, to find more effective ways to integrate them into the curricula. A complementary activity could be to invite experts from industry to teach soft skills to ICT students, giving students more practical perspectives while strengthening relationships between academia and industry.

\subsubsection*{Building the business case for sustainability} 
Sustainability and the SDGs provide enormous business opportunities for organisations~\cite{business2017better}. As one organisation in the interview study stated quite frankly ``\textit{Why we are so interested? [...] it's money.}'' In the mid-to-long term, companies can benefit from creating new systems to exploit these business opportunities. 

Therefore, IT professionals need to understand better what drives the businesses they work for, the opportunities that focus on sustainability open to businesses in general and to IT products and services in particular, and the threats faced by businesses and IT products causing harm to the environment and society. Understanding this might help them champion the idea of sustainability internally and justify it in terms of economic, environmental, and societal reasons. 
There exist a few methods and tools in Software Engineering to support building a business case, e.g. BeSusAF ~\cite{lammert2024bridging} and S-BGQM ~\cite{oyedeji2017sustainability}. In interviews with software practitioners, it has been argued that this is anyway not the core of a Software Engineer's work (building a business case) ~\cite{betz2022software}. However, a bridge between business cases, IT products and services, and sustainability impacts is needed to anticipate opportunities and threats for business, society, and the environment.  

This situation provides many opportunities for educators. For example, providing workshops on sustainability impacts for software engineers or developing example cases on sustainability impacts. Moreover, traditional practice-based courses such as capstones and hackathons could use these real-world challenges companies have and, as an outcome, provide possible solutions for companies to build business cases and minimise negative sustainability effects. In general, educators may collaborate with companies to increase practitioners' awareness of the possible sustainability impacts of their products and activities.

\subsubsection*{Legal Requirements and Standards} 
One of the pressures moving industry towards improved sustainability is the need to comply with legal requirements and adopt standards. In many jurisdictions, sustainability-related laws and regulations are becoming stricter and the penalties for non-compliance are increasing~\cite{Rajamani2021}. IT professionals need to be aware of how these may affect system requirements and how they may have an impact on business cases. Legal requirements may, for example, include carbon taxes, customer data privacy, mandatory sustainability reporting, Waste Electrical and Electronic Equipment (WEEE) disposal, and Environmental Impact Assessment for Data Centre construction proposals. The need to comply with these may have a direct impact on the requirements of a project. For example, Org. 14 reported that ``\textit{we’ve been working with our customers and see how EU regulations have evolved}''.
There is, in addition, a wide range of standards where compliance is not required by law but may be expected by customers and other stakeholders. For example, an increasing number of large businesses are requiring their suppliers to provide carbon footprint data (Scope 3 reporting)~\cite{GreenhouseGasProtocol2015} and many investors are expecting voluntary sustainability reporting complying with the Global Reporting Initiative or Sustainability Accounting Standards Board~\cite{Pizzi2023} Businesses can differentiate themselves in the marketplace by achieving certification to standards such as ISO14001 for Environmental Management Systems~\cite{BritishStandardsInstitute2015}. Some businesses may choose to change their legal foundation to benefit a wider range of stakeholders than just shareholders e.g. B Corps~\cite{BLab2018}. ICT professionals need education on the systematic impacts of such standards compliance and choices.  

\subsubsection*{Advocacy and lobbying}  
This topic raises the question of how impartial and neutral researchers and educators should be versus how much they are involved in advocacy and lobbying. We are in favour of taking an informed and principled stance while allowing discussion space for all perspectives on an issue. Some sustainability topics are grounded in established science, e.g. Climate Science~\cite{IntergovernmentalPanelonClimateChange2022}, and others are more determined by contestable, normative values, e.g. acceptable levels of inequality in society. Educators need to be explicit about their values and positionality. 
One should not wait for regulation to start acting on sustainability. Regulation is often a late follower of social trends and is highly influenced by them. Recent decades have witnessed a range of social and business movements towards sustainability, such as Corporate Social Responsibility, Fairtrade, Slow Food, UN Global Compact, Natural Capitalism, B-Corporations, etc. By contrast, technology often advances much faster and IT professionals may not adequately anticipate the sustainability impacts~\cite{Chatfield2017}.

The importance of offering expertise to policymakers was highlighted by Org. 9: ``\textit{a lot of policymakers don't have a clue on digitalisation matters, and because of that, they don't know what they're doing while writing the law.}". The IT professional bodies do engage in policy formation e.g. ACM, BCS, and educators can play a role in promoting and shaping movements such as the above~\cite{AssociationforComputingMachinery2024}. Universities should train future IT professionals to combine their technical and sustainability expertise to become strong advocates for sustainability. Therefore, curricula should also include tools for effective and positive advocacy in organisations, media and legislation, as well as lobbying.

\section{Under the Lens of our Sustainability in Computing Education Survey}\label{sec:takeaways}
As we critically reviewed the proposed topics we extracted from the interview material and the outcome of the literature review on sustainability in computing education~\cite{peters2024sustainability}, many of the same conclusions on computing education and open questions hold when reviewing what industry representatives argue. We offer similar questions as a result of the interview studies, and some tentative suggestions for ways of moving forward.

In the following, we provide a concise overview of our findings for each topic. 

The conceptualisations of sustainability were divergent and incoherent, both among the interviewees, but especially when contrasted with understandings of drivers of our current crises and inabilities to enact changes through research and education~\cite{Stoddard:2021aa}. The call for \textbf{core sustainability concepts} could be contrasted with the recommendation to make meaningful and substantive changes calibrated against drivers of our predicaments, and core topics related to equality, justice and power relations~\cite[Sec. 7.1.2]{peters2024sustainability}. However, the interviews presented a limited view of some of the symptoms of our interlinked crises, notably carbon emissions and climate change. As an open question, therefore, we ask how we may best move conversations forward to ensure a broader and more meaningful framing of topics needed, with industry professionals and educators alike.

In the literature review on computing education, recommendations for working with \textbf{values and ethics} address justice-centred computing education and methods for broader engagement with stakeholders marginalised or negatively affected by IT systems. As values thinking requires activities and explorations of the motivations and actions of learners themselves, and in contexts where they encounter stakeholders with different perspectives~\cite{Wiek:2015aa}, including values thinking might lead to disruptive types of learning if we allow students to explore values behind systems and question the underlying rationale for project proposals of IT systems in their capstone projects~\cite{duboc2020requirements}. For such disruptions to be possible and productive, they would have to not only include denouncing prevailing systems or practices but also open up pathways towards other forms of engaging with IT design. An open question here would be how to combine such critical approaches with systems design courses in a respectful and empathetic manner to gain broader acceptance for the need to critically review assumptions when designing IT systems.

\textbf{Systems thinking} requires a shift in frame and attention that would reposition reductionist computational thinking as subordinate to a systems perspective on how norms and power relations shape how we understand system boundaries and values of worth in IT design~\cite[Sec. 7.1.2]{peters2024sustainability}. An open question here would be how to empower educators and professionals trained in computational thinking to shift perspectives towards holistic, inclusive and critical analyses of systems, and how to include self-awareness in relation to limitations of knowledge.

\textbf{Sustainability impacts and measurements} are complementary, and at first glance perhaps seemingly opposed, to the holistic view brought about by systems thinking. An open question is whether a hands-on approach that delivers hard facts and measures progress is as important as a broad understanding of the intricate and complex reality. It might also have the benefit of motivating technically inclined students to embark on the journey and be more open to the systems thinking paradigm if they can also measure some parts of it. 

Understanding conditions for \textbf{technical sustainability} was called for as an important competence in the interviews. On the one hand, this is a recommendation aligned with what the literature review on education called for, by e.g. combining architecture assessment methods with longevity and broader concerns~\cite[Sec. 7.1.3]{peters2024sustainability}, but on the other, the longevity of a particular system might not be in the long-term interest of the social and ecological systems of which it is part. The literature review calls for caution, by saying that while ``understanding technical underpinnings and development trajectories of computing are relevant, reproducing or continuing on them might not be''~\cite[Section 8]{peters2024sustainability}. An open question therefore might be to best combine competencies to promote the longevity of designs, and the dispositions and abilities necessary to dismantle systems that are not in our long-term interest to maintain.

\textbf{Soft skills} are an integral part of what the interviews call for and could be interpreted as what the literature review concludes as important. However, the literature review does not use this particular term in recommendations for changes in curricula and the term has been criticised as pushing human-centred concerns outside of the purview of technical education, introducing an artificial distinction that makes it harder to see that all technology has social implications and is embedded in a social fabric of values~\cite{Berdanier:2022aa}. What the computing education literature review calls for instead is an explicit connection to research on norms, values, equality, justice and power, and to empower computing students to internalise stakeholders traditionally considered peripheral to design work. The types of skills necessary for wider stakeholder engagement may require methods of teaching and assessment that are currently some steps removed from the traditional types of teaching in computing. Training in empathetic communication, and having the ability to refrain from building destructive systems may challenge current notions of valid approaches to computing education. An open question then becomes how to provide training for computing educators, students and industry professionals alike, to make them comfortable to include notions of wider stakeholder dialogues, future thinking and empathetic listening to understand the social fabrics of which we are part.

\textbf{Building the business case for sustainability} is another topic not covered in the current computing education framework. Despite the perceived conflict between business interests and sustainability, industry stakeholders recognise potential business opportunities. Hence, we view this as an overlooked area that academia should explore further. Enhancing understanding of the business case for sustainability among students can help bridge this gap, and perhaps be added as a learning objective and possibly in the intersection between computing- and sustainability-specific topics.

For this to work, however, it is necessary to adopt a wide-boundary understanding of both symptoms and drivers of our current predicaments, and an open question remains about how to formulate general business responses that would be compatible with both short-term and long-term societal needs in light of them.

\textbf{Legal requirements and standards} are a core concern of businesses that might be too little acknowledged by academia (at least outside core fields such as management, although requirements engineering discusses legal issues and software quality courses discuss standards); in SWEBOK, for example, it is limited to one subtopic under the topic professionalism~\cite
[1.7 Legal Issues]{SWEBOK2014}.
Situating computing projects within the realm of real legal frameworks and requiring student projects to take legal implications seriously would probably help students understand the contexts in which industries operate. However, such legal frameworks on the one hand mandate better electronic waste management, but are still ineffective as they do not address the issues of design that make electronic equipment intrinsically hard to recycle~\cite{Shahabuddin:2023aa}. Legislation may even be counterproductive and mandate a maximisation of shareholder value at the expense of the values of others, or require public procurement at the lowest cost if no quantifiable attributes can separate a conscientious and a careless actor. An open question, therefore, would remain about how best to combine awareness of existing legal frameworks with a willingness to engage in sustainability and whether or not such work is supported by existing frameworks.  

Concerning \textbf{advocacy and lobbying}, as we imagine it, it goes far beyond simple awareness raising. As argued in \cite{peters2024sustainability}, we acknowledge that part of the scientific community feels the need for stronger lobbying on behalf of sustainability, and believe this to be also necessary~\cite{Gardner:2021ua}. If academia and business could align their efforts, such as by integrating business case studies and legal considerations into academic curricula, it would likely reduce resistance and opposition from influential industry stakeholders. This alignment may even foster cooperation and support towards achieving shared objectives. An open question that remains here is how such activism can be directed to calling for change within the business and education sectors themselves, and how to ensure that such activism can be taken seriously as benign, productive and empathetic but still firmly grounded in critical understanding of where we are and where we need to be.

\section{Conclusions} \label{sec:conclusions}
This paper analysed a set of sustainability-related topics as elicited from an interview study with computing professionals. The analysis is based on (i) our understanding of the results of interviews with IT professionals \cite{heldal2024sustainability}, (ii) our own expertise as researchers and educators engaged in teaching sustainability, and (iii) a comparison of those results to the recommendations from a recent literature review of sustainability in computing education~\cite{peters2024sustainability}. In particular, we provide open questions given how IT professionals reason and what the literature review called for. 

As we stated in \cite{heldal2024sustainability},
industry demands experts or engineers trained in specific sustainability topics. These demands combined with the recommendations based on the state of computing education~\cite{peters2024sustainability}, call for several shifts in the perception of how to reorient computing, such as we outlined in Section~\ref{sec:takeaways}.

Therefore, efforts are needed to change the status quo and propose a roadmap to incorporate sustainability learning objectives and courses in the computing education curricula. Here, computing educators and industry professionals may themselves be able to use a systems lens to understand which barriers to change exist, and how to employ leverage points for successful change in higher education in light of them~\cite{Blanco-Portela:2017aa,Leifler:2020ac}.

For instance, learning about both key competencies and learning activities conducive to promoting them~\cite{Lozano:2017aa,UNESCO:2017aa} and applying strategies for change agents in higher education~\cite{Moore:2005aa} would be necessary parts for a wider change process among computing educators.

Also, by openly asking the questions we pose in Section~\ref{sec:takeaways} to critical friends of computing as suggested in~\cite{Becker:2023aa}, and inviting them to help us define what computing should be in light of urgent needs for a new direction, computing can be a key leverage point towards just and sustainable futures~\cite{Creutzig:2022aa}. With these changes, we hope that education will be able to ``orient itself more toward emancipation, based on care, experimentation, critical thinking, and reflection''~\cite{peters2024sustainability}.

The agenda we foresee should align with European guidelines, such as the Green Deal education initiative and the EU recommendation (2022) to stimulate learning for the green transition and sustainable development.

\begin{acks}
The authors would like to thank all the interviewees who
took part in the study, and also:  NOVA LINCS (UIDB/04516/2020) with the financial support of FCT.IP; SFI SmartOcean NFR Project 309612/F40, by KISS (16DHBKI061) with the financial support of the Federal Ministry of Education and Research, Germany;  Catalan Government (Departament de Recerca i Universitats) for the grant 2021 SGR 01396 given to the HER group; VU Digital Sustainability Center (DiSC); Chalmers Area of Advance ICT for Sustainability project 37460087.
\end{acks}

\bibliographystyle{ACM-Reference-Format}
\bibliography{main}


\begin{thebibliography}{72}

%%% the Web address.

\ifx \showCODEN    \undefined \def \showCODEN     #1{\unskip}     \fi
\ifx \showDOI      \undefined \def \showDOI       #1{#1}\fi
\ifx \showISBNx    \undefined \def \showISBNx     #1{\unskip}     \fi
\ifx \showISBNxiii \undefined \def \showISBNxiii  #1{\unskip}     \fi
\ifx \showISSN     \undefined \def \showISSN      #1{\unskip}     \fi
\ifx \showLCCN     \undefined \def \showLCCN      #1{\unskip}     \fi
\ifx \shownote     \undefined \def \shownote      #1{#1}          \fi
\ifx \showarticletitle \undefined \def \showarticletitle #1{#1}   \fi
\ifx \showURL      \undefined \def \showURL       {\relax}        \fi
% invisible to TeX
\providecommand\bibfield[2]{#2}
\providecommand\bibinfo[2]{#2}
\providecommand\natexlab[1]{#1}
\providecommand\showeprint[2][]{arXiv:#2}

\bibitem[Alidoosti et~al\mbox{.}(2022a)]%
        {Alidoosti2022b}
\bibfield{author}{\bibinfo{person}{Raizeh Alidoosti}, \bibinfo{person}{Patricia Lago}, \bibinfo{person}{Eltjo Poort}, \bibinfo{person}{Maryam Razavian}, {and} \bibinfo{person}{Antony Tang}.} \bibinfo{year}{2022}\natexlab{a}.
\newblock \showarticletitle{{Incorporating Ethical Values into Software Architecture Design Practices}}. In \bibinfo{booktitle}{\emph{{19th International Conference on Software Architecture Companion ({ICSA-C})}}}. \bibinfo{publisher}{{IEEE}}, \bibinfo{pages}{124--127}.
\newblock


\bibitem[Alidoosti et~al\mbox{.}(2022b)]%
        {Alidoosti2022}
\bibfield{author}{\bibinfo{person}{Razieh Alidoosti}, \bibinfo{person}{Patricia Lago}, \bibinfo{person}{Maryam Razavian}, {and} \bibinfo{person}{Anthony Tang}.} \bibinfo{year}{2022}\natexlab{b}.
\newblock \bibinfo{booktitle}{\emph{{Ethics in Software Engineering: A Systematic Literature Review}}}.
\newblock \bibinfo{type}{{T}echnical {R}eport}. \bibinfo{institution}{Vrije Universiteit Amsterdam}.
\newblock
\urldef\tempurl%
\url{http://tiny.cc/bf74vz}
\showURL{%
\tempurl}


\bibitem[{Association for Computing Machinery}(2024)]%
        {AssociationforComputingMachinery2024}
\bibfield{author}{\bibinfo{person}{{Association for Computing Machinery}}.} \bibinfo{year}{2024}\natexlab{}.
\newblock \bibinfo{booktitle}{\emph{{ACM TechBriefs}}}.
\newblock \bibinfo{type}{{T}echnical {R}eport}. \bibinfo{address}{New York}.
\newblock
\urldef\tempurl%
\url{https://www.acm.org/public-policy/techbriefs}
\showURL{%
\tempurl}


\bibitem[{B Lab}(2018)]%
        {BLab2018}
\bibfield{author}{\bibinfo{person}{{B Lab}}.} \bibinfo{year}{2018}\natexlab{}.
\newblock \bibinfo{title}{{About B Corps}}.
\newblock
\newblock
\urldef\tempurl%
\url{https://bcorporation.net/about-b-corps}
\showURL{%
\tempurl}


\bibitem[Becker(2023)]%
        {Becker:2023aa}
\bibfield{author}{\bibinfo{person}{Christoph Becker}.} \bibinfo{year}{2023}\natexlab{}.
\newblock \bibinfo{booktitle}{\emph{Insolvent: How to Reorient Computing for Just Sustainability}}.
\newblock \bibinfo{publisher}{{MIT} Press}.
\newblock


\bibitem[Becker et~al\mbox{.}(2015b)]%
        {becker2015sustainability}
\bibfield{author}{\bibinfo{person}{Christoph Becker} {et~al\mbox{.}}} \bibinfo{year}{2015}\natexlab{b}.
\newblock \showarticletitle{Sustainability design and software: The {K}arlskrona manifesto}. In \bibinfo{booktitle}{\emph{2015 IEEE/ACM 37th IEEE International Conference on Software Engineering}}, Vol.~\bibinfo{volume}{2}. IEEE, \bibinfo{pages}{467--476}.
\newblock


\bibitem[Becker et~al\mbox{.}(2015a)]%
        {Karlskrona}
\bibfield{author}{\bibinfo{person}{Christoph Becker}, \bibinfo{person}{Ruzanna Chitchyan}, \bibinfo{person}{Leticia Duboc}, \bibinfo{person}{Steve Easterbrook}, \bibinfo{person}{Birgit Penzenstadler}, \bibinfo{person}{Norbert Seyff}, {and} \bibinfo{person}{Colin~C. Venters}.} \bibinfo{year}{2015}\natexlab{a}.
\newblock \showarticletitle{Sustainability Design and Software: The Karlskrona Manifesto}. In \bibinfo{booktitle}{\emph{Proceedings of the 37th International Conference on Software Engineering - Volume 2}} \emph{(\bibinfo{series}{ICSE '15})}. \bibinfo{publisher}{IEEE Press}, \bibinfo{address}{Piscataway, NJ, USA}, \bibinfo{pages}{467--476}.
\newblock
\urldef\tempurl%
\url{http://dl.acm.org/citation.cfm?id=2819009.2819082}
\showURL{%
\tempurl}


\bibitem[Berdanier(2022)]%
        {Berdanier:2022aa}
\bibfield{author}{\bibinfo{person}{Catherine~GP Berdanier}.} \bibinfo{year}{2022}\natexlab{}.
\newblock \showarticletitle{A hard stop to the term ``soft skills''}.
\newblock \bibinfo{journal}{\emph{Journal of Engineering Education}} \bibinfo{volume}{111}, \bibinfo{number}{1} (\bibinfo{year}{2022}), \bibinfo{pages}{14--18}.
\newblock


\bibitem[Betz et~al\mbox{.}(2022)]%
        {betz2022software}
\bibfield{author}{\bibinfo{person}{Stefanie Betz}, \bibinfo{person}{Dominic Lammert}, {and} \bibinfo{person}{Jari Porras}.} \bibinfo{year}{2022}\natexlab{}.
\newblock \showarticletitle{Software engineers in transition: Self-role attribution and awareness for sustainability}.
\newblock  (\bibinfo{year}{2022}).
\newblock


\bibitem[Bianchi et~al\mbox{.}(2022)]%
        {RePEc:ipt:iptwpa:jrc128040}
\bibfield{author}{\bibinfo{person}{Guia Bianchi}, \bibinfo{person}{Ulrike Pisiotis}, {and} \bibinfo{person}{Marcelino Cabrera~Giraldez}.} \bibinfo{year}{2022}\natexlab{}.
\newblock \bibinfo{booktitle}{\emph{GreenComp The European sustainability competence framework}}.
\newblock \bibinfo{type}{JRC Research Reports} JRC128040. \bibinfo{institution}{Joint Research Centre (Seville site)}.
\newblock
\urldef\tempurl%
\url{https://EconPapers.repec.org/RePEc:ipt:iptwpa:jrc128040}
\showURL{%
\tempurl}


\bibitem[Blanco-Portela et~al\mbox{.}(2017)]%
        {Blanco-Portela:2017aa}
\bibfield{author}{\bibinfo{person}{Norka Blanco-Portela}, \bibinfo{person}{Javier Benayas}, \bibinfo{person}{Luis~R. Pertierrac}, {and} \bibinfo{person}{Rodrigo Lozano}.} \bibinfo{year}{2017}\natexlab{}.
\newblock \showarticletitle{Towards the integration of sustainability in Higher Education Institutions: A review of drivers of and barriers to organisational change and their comparison against those found of companies}.
\newblock \bibinfo{journal}{\emph{Journal of Cleaner Production}}  \bibinfo{volume}{166} (\bibinfo{year}{2017}), \bibinfo{pages}{563--578}.
\newblock


\bibitem[Bourque and Fairley(2014)]%
        {SWEBOK2014}
\bibfield{editor}{\bibinfo{person}{Pierre Bourque} {and} \bibinfo{person}{Richard~E. Fairley}} (Eds.). \bibinfo{year}{2014}\natexlab{}.
\newblock \bibinfo{booktitle}{\emph{{SWEBOK}: Guide to the Software Engineering Body of Knowledge} (\bibinfo{edition}{version 3.0} ed.)}.
\newblock \bibinfo{publisher}{IEEE Computer Society}, \bibinfo{address}{Los Alamitos, CA}.
\newblock
\showISBNx{978-0-7695-5166-1}
\urldef\tempurl%
\url{http://www.swebok.org/}
\showURL{%
\tempurl}


\bibitem[Bremer et~al\mbox{.}(2023)]%
        {bremer23:assessing}
\bibfield{author}{\bibinfo{person}{Christina Bremer} {et~al\mbox{.}}} \bibinfo{year}{2023}\natexlab{}.
\newblock \bibinfo{booktitle}{\emph{{Assessing Energy and Climate Effects of Digitalization: Methodological Challenges and Key Recommendations}}}.
\newblock \bibinfo{type}{{T}echnical {R}eport}. \bibinfo{institution}{nDEE Framing Paper Series}.
\newblock
\urldef\tempurl%
\url{https://doi.org/10.2139/ssrn.4459526}
\showDOI{\tempurl}


\bibitem[{British Standards Institute}(2015)]%
        {BritishStandardsInstitute2015}
\bibfield{author}{\bibinfo{person}{{British Standards Institute}}.} \bibinfo{year}{2015}\natexlab{}.
\newblock \bibinfo{title}{{BS EN ISO 14001:2015: Environmental management systems. Requirements with guidance for use}}.
\newblock , \bibinfo{numpages}{48}~pages.
\newblock
\showISBNx{0580826112;9780580826115;}
\urldef\tempurl%
\url{https://knowledge.bsigroup.com/products/environmental-management-systems-requirements-with-guidance-for-use?version=tracked}
\showURL{%
\tempurl}


\bibitem[Burnett et~al\mbox{.}(2016)]%
        {burnett2016gendermag}
\bibfield{author}{\bibinfo{person}{Margaret Burnett} {et~al\mbox{.}}} \bibinfo{year}{2016}\natexlab{}.
\newblock \showarticletitle{GenderMag: A method for evaluating software's gender inclusiveness}.
\newblock \bibinfo{journal}{\emph{Interacting with Computers}} \bibinfo{volume}{28}, \bibinfo{number}{6} (\bibinfo{year}{2016}), \bibinfo{pages}{760--787}.
\newblock


\bibitem[Chalmers(2019)]%
        {chalmers2019sdg}
\bibfield{author}{\bibinfo{person}{GMV Chalmers}.} \bibinfo{year}{2019}\natexlab{}.
\newblock \showarticletitle{The SDG Impact Assessment Tool-a free online tool for self-assessments of impacts on Agenda 2030}.
\newblock \bibinfo{journal}{\emph{Policy}}  \bibinfo{volume}{1} (\bibinfo{year}{2019}), \bibinfo{pages}{150--167}.
\newblock


\bibitem[Chatfield et~al\mbox{.}(2017)]%
        {Chatfield2017}
\bibfield{author}{\bibinfo{person}{Kate Chatfield}, \bibinfo{person}{Elisabetta Borsella}, \bibinfo{person}{Elvio Mantovani}, \bibinfo{person}{Andrea Porcari}, {and} \bibinfo{person}{Bernd~Carsten Stahl}.} \bibinfo{year}{2017}\natexlab{}.
\newblock \showarticletitle{{An investigation into risk perception in the ICT industry as a core component of responsible research and innovation}}.
\newblock \bibinfo{journal}{\emph{Sustainability (Basel, Switzerland)}} \bibinfo{volume}{9}, \bibinfo{number}{8} (\bibinfo{year}{2017}), \bibinfo{pages}{1424}.
\newblock
\showISSN{2071-1050}
\urldef\tempurl%
\url{https://doi.org/10.3390/su9081424}
\showDOI{\tempurl}


\bibitem[Chitchyan et~al\mbox{.}(2016)]%
        {chitchyan2016}
\bibfield{author}{\bibinfo{person}{Ruzanna Chitchyan} {et~al\mbox{.}}} \bibinfo{year}{2016}\natexlab{}.
\newblock \showarticletitle{Sustainability Design in Requirements Engineering: State of Practice}. In \bibinfo{booktitle}{\emph{Proceedings of the 38th International Conference on Software Engineering Companion}}. \bibinfo{pages}{533–542}.
\newblock


\bibitem[Creutzig et~al\mbox{.}(2022)]%
        {Creutzig:2022aa}
\bibfield{author}{\bibinfo{person}{Felix Creutzig} {et~al\mbox{.}}} \bibinfo{year}{2022}\natexlab{}.
\newblock \showarticletitle{Digitalization and the Anthropocene}.
\newblock \bibinfo{journal}{\emph{Annual review of environment and resources}}  \bibinfo{volume}{47} (\bibinfo{year}{2022}), \bibinfo{pages}{479--509}.
\newblock


\bibitem[Duboc et~al\mbox{.}(2020a)]%
        {duboc2020requirements}
\bibfield{author}{\bibinfo{person}{Leticia Duboc} {et~al\mbox{.}}} \bibinfo{year}{2020}\natexlab{a}.
\newblock \showarticletitle{Requirements engineering for sustainability: an awareness framework for designing software systems for a better tomorrow}.
\newblock \bibinfo{journal}{\emph{Requirements Engineering}} \bibinfo{volume}{25}, \bibinfo{number}{4} (\bibinfo{year}{2020}), \bibinfo{pages}{469--492}.
\newblock


\bibitem[Duboc et~al\mbox{.}(2020b)]%
        {Duboc:2020aa}
\bibfield{author}{\bibinfo{person}{Leticia Duboc}, \bibinfo{person}{Curtis McCord}, \bibinfo{person}{Christoph Becker}, {and} \bibinfo{person}{Syed~Ishtiaque Ahmed}.} \bibinfo{year}{2020}\natexlab{b}.
\newblock \showarticletitle{Critical Requirements Engineering in Practice}.
\newblock \bibinfo{journal}{\emph{{IEEE} Software}} (\bibinfo{date}{February} \bibinfo{year}{2020}).
\newblock


\bibitem[Fiselier and Longhurst(2018)]%
        {fiselier2018critical}
\bibfield{author}{\bibinfo{person}{Evelien~S Fiselier} {and} \bibinfo{person}{James~WS Longhurst}.} \bibinfo{year}{2018}\natexlab{}.
\newblock \showarticletitle{A critical evaluation of the representation of the QAA and HEA guidance on ESD in public web environments of UK higher education institutions}.
\newblock \bibinfo{journal}{\emph{Implementing sustainability in the curriculum of universities: Approaches, methods and projects}} (\bibinfo{year}{2018}), \bibinfo{pages}{223--246}.
\newblock


\bibitem[Gardner et~al\mbox{.}(2021)]%
        {Gardner:2021ua}
\bibfield{author}{\bibinfo{person}{Charlie~J. Gardner}, \bibinfo{person}{Aaron Thierry}, \bibinfo{person}{William Rowlandson}, {and} \bibinfo{person}{Julia~K. Steinberger}.} \bibinfo{year}{2021}\natexlab{}.
\newblock \showarticletitle{From Publications to Public Actions: The Role of Universities in Facilitating Academic Advocacy and Activism in the Climate and Ecological Emergency}.
\newblock \bibinfo{journal}{\emph{Frontiers in Sustainability}}  \bibinfo{volume}{2} (\bibinfo{year}{2021}).
\newblock


\bibitem[Gharbi et~al\mbox{.}(2019)]%
        {Gharbi2019}
\bibfield{author}{\bibinfo{person}{Mahbouba Gharbi}, \bibinfo{person}{Arne Koschel}, {and} \bibinfo{person}{Andreas Rausch}.} \bibinfo{year}{2019}\natexlab{}.
\newblock \bibinfo{booktitle}{\emph{Software Architecture Fundamentals: A Study Guide for the Certified Professional for Software Architecture{\textregistered}--Foundation Level--iSAQB compliant}}.
\newblock \bibinfo{publisher}{dpunkt. verlag}.
\newblock


\bibitem[Glasser and Hirsh(2016)]%
        {glasser2016toward}
\bibfield{author}{\bibinfo{person}{Harold Glasser} {and} \bibinfo{person}{Jamie Hirsh}.} \bibinfo{year}{2016}\natexlab{}.
\newblock \showarticletitle{Toward the development of robust learning for sustainability core competencies}.
\newblock \bibinfo{journal}{\emph{Sustainability: The Journal of Record}} \bibinfo{volume}{9}, \bibinfo{number}{3} (\bibinfo{year}{2016}), \bibinfo{pages}{121--134}.
\newblock


\bibitem[Gotterbarn et~al\mbox{.}(2017)]%
        {gotterbarn2017acm}
\bibfield{author}{\bibinfo{person}{Don Gotterbarn}, \bibinfo{person}{Amy Bruckman}, \bibinfo{person}{Catherine Flick}, \bibinfo{person}{Keith Miller}, {and} \bibinfo{person}{Marty~J Wolf}.} \bibinfo{year}{2017}\natexlab{}.
\newblock \showarticletitle{ACM code of ethics: a guide for positive action}.
\newblock \bibinfo{journal}{\emph{Commun. ACM}} \bibinfo{volume}{61}, \bibinfo{number}{1} (\bibinfo{year}{2017}), \bibinfo{pages}{121--128}.
\newblock


\bibitem[Gotterbarn et~al\mbox{.}(1997)]%
        {gotterbarn1997software}
\bibfield{author}{\bibinfo{person}{Don Gotterbarn}, \bibinfo{person}{Keith Miller}, {and} \bibinfo{person}{Simon Rogerson}.} \bibinfo{year}{1997}\natexlab{}.
\newblock \showarticletitle{Software engineering code of ethics}.
\newblock \bibinfo{journal}{\emph{Commun. ACM}} \bibinfo{volume}{40}, \bibinfo{number}{11} (\bibinfo{year}{1997}), \bibinfo{pages}{110--118}.
\newblock


\bibitem[{Greenhouse Gas Protocol}(2015)]%
        {GreenhouseGasProtocol2015}
\bibfield{author}{\bibinfo{person}{{Greenhouse Gas Protocol}}.} \bibinfo{year}{2015}\natexlab{}.
\newblock \bibinfo{title}{{Corporate Standard}}.
\newblock
\newblock
\urldef\tempurl%
\url{http://www.ghgprotocol.org/corporate-standard}
\showURL{%
\tempurl}


\bibitem[Gregory(1979)]%
        {Gregory:1979aa}
\bibfield{author}{\bibinfo{person}{Bateson Gregory}.} \bibinfo{year}{1979}\natexlab{}.
\newblock \showarticletitle{Mind and nature: A necessary unity}.
\newblock \bibinfo{journal}{\emph{New York: Dutton}} (\bibinfo{year}{1979}).
\newblock


\bibitem[Groher and Weinreich(2017)]%
        {Groher2017}
\bibfield{author}{\bibinfo{person}{Iris Groher} {and} \bibinfo{person}{Rainer Weinreich}.} \bibinfo{year}{2017}\natexlab{}.
\newblock \showarticletitle{An Interview Study on Sustainability Concerns in Software Development Projects}. In \bibinfo{booktitle}{\emph{2017 43rd Euromicro Conference on Software Engineering and Advanced Applications (SEAA)}}. \bibinfo{pages}{350--358}.
\newblock
\urldef\tempurl%
\url{https://doi.org/10.1109/SEAA.2017.70}
\showDOI{\tempurl}


\bibitem[Heldal et~al\mbox{.}(2024)]%
        {heldal2024sustainability}
\bibfield{author}{\bibinfo{person}{Rogardt Heldal} {et~al\mbox{.}}} \bibinfo{year}{2024}\natexlab{}.
\newblock \showarticletitle{Sustainability competencies and skills in software engineering: An industry perspective}.
\newblock \bibinfo{journal}{\emph{Journal of Systems and Software}} (\bibinfo{year}{2024}), \bibinfo{pages}{111978}.
\newblock


\bibitem[Hilty and Aebischer(2015)]%
        {hilty2015ict}
\bibfield{author}{\bibinfo{person}{Lorenz~M Hilty} {and} \bibinfo{person}{Bernard Aebischer}.} \bibinfo{year}{2015}\natexlab{}.
\newblock \showarticletitle{{ICT} for sustainability: An emerging research field}.
\newblock In \bibinfo{booktitle}{\emph{ICT innovations for Sustainability}}. \bibinfo{publisher}{Springer}, \bibinfo{pages}{3--36}.
\newblock


\bibitem[{Intergovernmental Panel on Climate Change}(2022)]%
        {IntergovernmentalPanelonClimateChange2022}
\bibfield{author}{\bibinfo{person}{{Intergovernmental Panel on Climate Change}}.} \bibinfo{year}{2022}\natexlab{}.
\newblock \bibinfo{title}{{Sixth Assessment Report}}.
\newblock
\newblock
\urldef\tempurl%
\url{https://www.ipcc.ch/assessment-report/ar6/}
\showURL{%
\tempurl}


\bibitem[Jacobson(2018)]%
        {Jacobson2018}
\bibfield{author}{\bibinfo{person}{Ivar Jacobson}.} \bibinfo{year}{2018}\natexlab{}.
\newblock \bibinfo{title}{{50 years of SE, so now what?}}
\newblock \bibinfo{howpublished}{40th International Conference on Software Engineering (ICSE)}.
\newblock


\bibitem[Kimmerer(2013)]%
        {Kimmerer:2013aa}
\bibfield{author}{\bibinfo{person}{Robin Kimmerer}.} \bibinfo{year}{2013}\natexlab{}.
\newblock \bibinfo{booktitle}{\emph{Braiding sweetgrass: Indigenous wisdom, scientific knowledge and the teachings of plants}}.
\newblock \bibinfo{publisher}{Milkweed editions}.
\newblock


\bibitem[Kumar and Buyya(2012)]%
        {kumar2012green}
\bibfield{author}{\bibinfo{person}{Saurabh Kumar} {and} \bibinfo{person}{Rajkumar Buyya}.} \bibinfo{year}{2012}\natexlab{}.
\newblock \showarticletitle{Green cloud computing and environmental sustainability}.
\newblock \bibinfo{journal}{\emph{Harnessing green IT: principles and practices}} (\bibinfo{year}{2012}), \bibinfo{pages}{315--339}.
\newblock


\bibitem[Lago(2019)]%
        {Lago2019}
\bibfield{author}{\bibinfo{person}{P Lago}.} \bibinfo{year}{2019}\natexlab{}.
\newblock \showarticletitle{{Architecture Design Decision Maps for Software Sustainability}}. In \bibinfo{booktitle}{\emph{{41st International Conference on Software Engineering: Software Engineering in Society ({ICSE-SEIS})}}}. {IEEE/ACM}, \bibinfo{pages}{61--64}.
\newblock


\bibitem[Lago and Condori-Fernandez(2022)]%
        {SAFToolkit}
\bibfield{author}{\bibinfo{person}{Patricia Lago} {and} \bibinfo{person}{Nelly Condori-Fernandez}.} \bibinfo{year}{2022}\natexlab{}.
\newblock \bibinfo{title}{{The Sustainability Assessment Framework ({SAF}) Toolkit: Instruments to help Sustainability-driven Software Architecture Design Decision Making}}.
\newblock
\newblock
\urldef\tempurl%
\url{https://github.com/S2-group/SAF-Toolkit}
\showURL{%
\tempurl}


\bibitem[Lammert(2024)]%
        {lammert2024bridging}
\bibfield{author}{\bibinfo{person}{Dominic Lammert}.} \bibinfo{year}{2024}\natexlab{}.
\newblock \showarticletitle{Bridging academic software sustainability design with corporate business planning}.
\newblock  (\bibinfo{year}{2024}).
\newblock


\bibitem[Leifler and Dahlin(2020)]%
        {Leifler:2020ac}
\bibfield{author}{\bibinfo{person}{Ola Leifler} {and} \bibinfo{person}{Jon-Erik Dahlin}.} \bibinfo{year}{2020}\natexlab{}.
\newblock \showarticletitle{Curriculum integration of sustainability in engineering education--a national study of programme director perspectives}.
\newblock \bibinfo{journal}{\emph{International Journal of Sustainability in Higher Education}} \bibinfo{volume}{21}, \bibinfo{number}{5} (\bibinfo{year}{2020}), \bibinfo{pages}{877--894}.
\newblock


\bibitem[Lilienthal(2019)]%
        {Lilienthal2019}
\bibfield{author}{\bibinfo{person}{Carola Lilienthal}.} \bibinfo{year}{2019}\natexlab{}.
\newblock \bibinfo{booktitle}{\emph{Sustainable software architecture: analyze and reduce technical debt} (\bibinfo{edition}{1st} ed.)}.
\newblock


\bibitem[Lima et~al\mbox{.}(2022)]%
        {lima2022marine}
\bibfield{author}{\bibinfo{person}{Keila Lima} {et~al\mbox{.}}} \bibinfo{year}{2022}\natexlab{}.
\newblock \showarticletitle{Marine data sharing: Challenges, technology drivers and quality attributes}. In \bibinfo{booktitle}{\emph{International Conference on Product-Focused Software Process Improvement}}. Springer, \bibinfo{pages}{124--140}.
\newblock


\bibitem[Linn{\'e}r and Wibeck(2021)]%
        {Linner:2021aa}
\bibfield{author}{\bibinfo{person}{Bj{\"o}rn-Ola Linn{\'e}r} {and} \bibinfo{person}{Victoria Wibeck}.} \bibinfo{year}{2021}\natexlab{}.
\newblock \showarticletitle{Drivers of sustainability transformations: leverage points, contexts and conjunctures}.
\newblock \bibinfo{journal}{\emph{Sustainability Science}} \bibinfo{volume}{16}, \bibinfo{number}{3} (\bibinfo{year}{2021}), \bibinfo{pages}{889--900}.
\newblock


\bibitem[Lozano et~al\mbox{.}(2017)]%
        {Lozano:2017aa}
\bibfield{author}{\bibinfo{person}{Rodrigo Lozano}, \bibinfo{person}{Michelle~Y. Merrill}, \bibinfo{person}{Kaisu Sammalisto}, \bibinfo{person}{Kim Ceulemans}, {and} \bibinfo{person}{Francisco~J. Lozano}.} \bibinfo{year}{2017}\natexlab{}.
\newblock \showarticletitle{Connecting Competences and Pedagogical Approaches for Sustainable Development in Higher Education: A Literature Review and Framework Proposal}.
\newblock \bibinfo{journal}{\emph{Sustainability}} \bibinfo{volume}{9}, \bibinfo{number}{10} (\bibinfo{year}{2017}).
\newblock


\bibitem[Marin-Zapata et~al\mbox{.}(2022)]%
        {Marin-ZapataSaraIsabel2022Ssdw}
\bibfield{author}{\bibinfo{person}{Sara~Isabel Marin-Zapata}, \bibinfo{person}{Juan~Pablo Román-Calderón}, \bibinfo{person}{Cristina Robledo-Ardila}, {and} \bibinfo{person}{Maria~Alejandra Jaramillo-Serna}.} \bibinfo{year}{2022}\natexlab{}.
\newblock \showarticletitle{Soft skills, do we know what we are talking about?}
\newblock \bibinfo{journal}{\emph{Review of managerial science}} \bibinfo{volume}{16}, \bibinfo{number}{4} (\bibinfo{year}{2022}), \bibinfo{pages}{969--1000}.
\newblock
\showISSN{1863-6683}


\bibitem[McPhearson et~al\mbox{.}(2021)]%
        {McPhearson:2021aa}
\bibfield{author}{\bibinfo{person}{Timon McPhearson} {et~al\mbox{.}}} \bibinfo{year}{2021}\natexlab{}.
\newblock \showarticletitle{Radical changes are needed for transformations to a good Anthropocene}.
\newblock \bibinfo{journal}{\emph{Npj urban sustainability}} \bibinfo{volume}{1}, \bibinfo{number}{1} (\bibinfo{year}{2021}), \bibinfo{pages}{5}.
\newblock


\bibitem[Moore(2005)]%
        {Moore:2005aa}
\bibfield{author}{\bibinfo{person}{Janet Moore}.} \bibinfo{year}{2005}\natexlab{}.
\newblock \showarticletitle{Seven recommendations for creating sustainability education at the university level: A guide for change agents}.
\newblock \bibinfo{journal}{\emph{International Journal of Sustainability in Higher Education}} \bibinfo{volume}{6}, \bibinfo{number}{4} (\bibinfo{year}{2005}), \bibinfo{pages}{326--339}.
\newblock


\bibitem[Nguyen et~al\mbox{.}(2023)]%
        {IoTJournal}
\bibfield{author}{\bibinfo{person}{Ngoc-Thanh Nguyen} {et~al\mbox{.}}} \bibinfo{year}{2023}\natexlab{}.
\newblock \showarticletitle{Engineering Challenges of Stationary Wireless Smart Ocean Observation Systems}.
\newblock \bibinfo{journal}{\emph{IEEE Internet of Things Journal}} \bibinfo{volume}{10}, \bibinfo{number}{16} (\bibinfo{year}{2023}), \bibinfo{pages}{14712--14724}.
\newblock


\bibitem[Olmstead et~al\mbox{.}(1974)]%
        {olmstead1974research}
\bibfield{author}{\bibinfo{person}{Joseph~A Olmstead} {et~al\mbox{.}}} \bibinfo{year}{1974}\natexlab{}.
\newblock \showarticletitle{Research on Utilization of Assessment Results and Methods. Final Technical Report.}
\newblock  (\bibinfo{year}{1974}).
\newblock


\bibitem[Oppenheim et~al\mbox{.}(2017)]%
        {business2017better}
\bibfield{author}{\bibinfo{person}{Jeremy Oppenheim} {et~al\mbox{.}}} \bibinfo{year}{2017}\natexlab{}.
\newblock \showarticletitle{Better business, better world}.
\newblock \bibinfo{journal}{\emph{Business \& Sustainable Development Commission}} (\bibinfo{year}{2017}).
\newblock


\bibitem[Oyedeji et~al\mbox{.}(2017)]%
        {oyedeji2017sustainability}
\bibfield{author}{\bibinfo{person}{Shola Oyedeji}, \bibinfo{person}{Ahmed Seffah}, {and} \bibinfo{person}{Birgit Penzenstadler}.} \bibinfo{year}{2017}\natexlab{}.
\newblock \showarticletitle{Sustainability quantification in requirements informing design}.
\newblock \bibinfo{journal}{\emph{6th Int. Work. Requir. Eng. Sustain. Syst}}  \bibinfo{volume}{1} (\bibinfo{year}{2017}).
\newblock


\bibitem[Penzenstadler(2014)]%
        {penzenstadler2014infusing}
\bibfield{author}{\bibinfo{person}{Birgit Penzenstadler}.} \bibinfo{year}{2014}\natexlab{}.
\newblock \showarticletitle{Infusing Green: Requirements Engineering for Green In and Through Software Systems.}. In \bibinfo{booktitle}{\emph{RE4SuSy@ RE}}. \bibinfo{pages}{44--53}.
\newblock


\bibitem[Peters et~al\mbox{.}(2024)]%
        {peters2024sustainability}
\bibfield{author}{\bibinfo{person}{Anne-Kathrin Peters} {et~al\mbox{.}}} \bibinfo{year}{2024}\natexlab{}.
\newblock \showarticletitle{Sustainability in computing education: A systematic literature review}.
\newblock \bibinfo{journal}{\emph{ACM Transactions on Computing Education}} \bibinfo{volume}{24}, \bibinfo{number}{1} (\bibinfo{year}{2024}), \bibinfo{pages}{1--53}.
\newblock


\bibitem[Pihkala(2022)]%
        {Pihkala:2022aa}
\bibfield{author}{\bibinfo{person}{Panu Pihkala}.} \bibinfo{year}{2022}\natexlab{}.
\newblock \showarticletitle{The Process of Eco-Anxiety and Ecological Grief: A Narrative Review and a New Proposal}.
\newblock \bibinfo{journal}{\emph{Sustainability}} \bibinfo{volume}{14}, \bibinfo{number}{24} (\bibinfo{year}{2022}).
\newblock


\bibitem[Pizzi et~al\mbox{.}(2023)]%
        {Pizzi2023}
\bibfield{author}{\bibinfo{person}{Simone Pizzi}, \bibinfo{person}{Salvatore Principale}, {and} \bibinfo{person}{Elbano de Nuccio}.} \bibinfo{year}{2023}\natexlab{}.
\newblock \showarticletitle{{Material sustainability information and reporting standards. Exploring the differences between GRI and SASB}}.
\newblock \bibinfo{journal}{\emph{Meditari accountancy research}} \bibinfo{volume}{31}, \bibinfo{number}{6} (\bibinfo{year}{2023}), \bibinfo{pages}{1654--1674}.
\newblock
\showISSN{2049-372X}
\urldef\tempurl%
\url{https://doi.org/10.1108/MEDAR-11-2021-1486}
\showDOI{\tempurl}


\bibitem[{Quality Assurance Agency for Higher Education} and {Advance HE}(2021)]%
        {QualityAssuranceAgencyforHigherEducation2021}
\bibfield{author}{\bibinfo{person}{{Quality Assurance Agency for Higher Education}} {and} \bibinfo{person}{{Advance HE}}.} \bibinfo{year}{2021}\natexlab{}.
\newblock \bibinfo{booktitle}{\emph{{Education for Sustainable Development Guidance}}}.
\newblock \bibinfo{type}{{T}echnical {R}eport}. \bibinfo{institution}{QAA and Advance HE}, \bibinfo{address}{Gloucester}. \bibinfo{pages}{1--51} pages.
\newblock
\urldef\tempurl%
\url{https://membershipresources.qaa.ac.uk/s/article/Education-for-Sustainable-Development-Guidance}
\showURL{%
\tempurl}


\bibitem[Rajamani and Peel(2021)]%
        {Rajamani2021}
\bibfield{author}{\bibinfo{person}{Lavanya Rajamani} {and} \bibinfo{person}{Jacqueline Peel}.} \bibinfo{year}{2021}\natexlab{}.
\newblock \showarticletitle{{Reflections on a decade of change in international environmental law}}.
\newblock \bibinfo{journal}{\emph{Cambridge International Law Journal}} \bibinfo{volume}{10}, \bibinfo{number}{1} (\bibinfo{year}{2021}), \bibinfo{pages}{6--31}.
\newblock
\showISSN{2398-9173}
\urldef\tempurl%
\url{https://doi.org/10.4337/cilj.2021.01.01}
\showDOI{\tempurl}


\bibitem[Ramage and Shipp(2020)]%
        {Ramage:2020aa}
\bibfield{author}{\bibinfo{person}{Magnus Ramage} {and} \bibinfo{person}{Karen Shipp}.} \bibinfo{year}{2020}\natexlab{}.
\newblock \bibinfo{booktitle}{\emph{Systems Thinkers} (\bibinfo{edition}{second edition} ed.)}.
\newblock \bibinfo{publisher}{Springer}.
\newblock


\bibitem[Raworth(2017)]%
        {Raworth2017}
\bibfield{author}{\bibinfo{person}{Kate Raworth}.} \bibinfo{year}{2017}\natexlab{}.
\newblock \bibinfo{booktitle}{\emph{{Doughnut economics: seven ways to think like a 21st century economist}}}.
\newblock \bibinfo{publisher}{Chelsea Green Publishing}, \bibinfo{address}{White River Junction, Vermont}.
\newblock
\showISBNx{1847941370;9781603586740;1603586741;9781847941374;}


\bibitem[Richardson et~al\mbox{.}(2023)]%
        {Richardson2023}
\bibfield{author}{\bibinfo{person}{Katherine Richardson}, \bibinfo{person}{Will Steffen}, \bibinfo{person}{Wolfgang Lucht}, \bibinfo{person}{J{\o}rgen Bendtsen}, \bibinfo{person}{Sarah~E Cornell}, \bibinfo{person}{Jonathan~F Donges}, \bibinfo{person}{Markus Dr{\"{u}}ke}, \bibinfo{person}{Ingo Fetzer}, \bibinfo{person}{Govindasamy Bala}, \bibinfo{person}{Werner von Bloh}, \bibinfo{person}{Georg Feulner}, \bibinfo{person}{Stephanie Fiedler}, \bibinfo{person}{Dieter Gerten}, \bibinfo{person}{Tom Gleeson}, \bibinfo{person}{Matthias Hofmann}, \bibinfo{person}{Willem Huiskamp}, \bibinfo{person}{Matti Kummu}, \bibinfo{person}{Chinchu Mohan}, \bibinfo{person}{David Nogu{\'{e}}s-Bravo}, \bibinfo{person}{Stefan Petri}, \bibinfo{person}{Miina Porkka}, \bibinfo{person}{Stefan Rahmstorf}, \bibinfo{person}{Sibyll Schaphoff}, \bibinfo{person}{Kirsten Thonicke}, \bibinfo{person}{Arne Tobian}, \bibinfo{person}{Vili Virkki}, \bibinfo{person}{Lan Wang-Erlandsson}, \bibinfo{person}{Lisa Weber}, {and} \bibinfo{person}{Johan
  Rockstr{\"{o}}m}.} \bibinfo{year}{2023}\natexlab{}.
\newblock \showarticletitle{{Earth beyond six of nine planetary boundaries}}.
\newblock \bibinfo{journal}{\emph{Science Advances}} \bibinfo{volume}{9}, \bibinfo{number}{37} (\bibinfo{date}{sep} \bibinfo{year}{2023}), \bibinfo{pages}{eadh2458}.
\newblock
\urldef\tempurl%
\url{https://doi.org/10.1126/sciadv.adh2458}
\showDOI{\tempurl}


\bibitem[Schwartz(2007)]%
        {schwartz2007basic}
\bibfield{author}{\bibinfo{person}{Shalom~H Schwartz}.} \bibinfo{year}{2007}\natexlab{}.
\newblock \showarticletitle{Basic human values: Theory, measurement, and applications}.
\newblock \bibinfo{journal}{\emph{Revue fran{\c{c}}aise de sociologie}} \bibinfo{volume}{47}, \bibinfo{number}{4} (\bibinfo{year}{2007}), \bibinfo{pages}{929}.
\newblock


\bibitem[Seacord et~al\mbox{.}(2003)]%
        {Seacord2003}
\bibfield{author}{\bibinfo{person}{R.C. Seacord}, \bibinfo{person}{J. Elm}, \bibinfo{person}{W. Goethert}, \bibinfo{person}{G.A. Lewis}, \bibinfo{person}{D. Plakosh}, \bibinfo{person}{J. Robert}, \bibinfo{person}{L. Wrage}, {and} \bibinfo{person}{M. Lindvall}.} \bibinfo{year}{2003}\natexlab{}.
\newblock \showarticletitle{Measuring software sustainability}. In \bibinfo{booktitle}{\emph{International Conference on Software Maintenance, 2003. ICSM 2003. Proceedings.}} \bibinfo{pages}{450--459}.
\newblock
\urldef\tempurl%
\url{https://doi.org/10.1109/ICSM.2003.1235455}
\showDOI{\tempurl}


\bibitem[Shahabuddin et~al\mbox{.}(2023)]%
        {Shahabuddin:2023aa}
\bibfield{author}{\bibinfo{person}{M Shahabuddin}, \bibinfo{person}{M~Nur Uddin}, \bibinfo{person}{JI Chowdhury}, \bibinfo{person}{SF Ahmed}, \bibinfo{person}{MN Uddin}, \bibinfo{person}{M Mofijur}, {and} \bibinfo{person}{MA Uddin}.} \bibinfo{year}{2023}\natexlab{}.
\newblock \showarticletitle{A review of the recent development, challenges, and opportunities of electronic waste (e-waste)}.
\newblock \bibinfo{journal}{\emph{International Journal of Environmental Science and Technology}} \bibinfo{volume}{20}, \bibinfo{number}{4} (\bibinfo{year}{2023}), \bibinfo{pages}{4513--4520}.
\newblock


\bibitem[Shearman(1990)]%
        {shearman1990meaning}
\bibfield{author}{\bibinfo{person}{Richard Shearman}.} \bibinfo{year}{1990}\natexlab{}.
\newblock \showarticletitle{The meaning and ethics of sustainability}.
\newblock \bibinfo{journal}{\emph{Environmental management}}  \bibinfo{volume}{14} (\bibinfo{year}{1990}), \bibinfo{pages}{1--8}.
\newblock


\bibitem[St{\aa}lne and Greca(2022)]%
        {Stalne:2022aa}
\bibfield{author}{\bibinfo{person}{Kristian St{\aa}lne} {and} \bibinfo{person}{Stefanie Greca}.} \bibinfo{year}{2022}\natexlab{}.
\newblock \bibinfo{title}{Inner Development Goals}.
\newblock \bibinfo{howpublished}{https://www.innerdevelopmentgoals.org/resources}.
\newblock


\bibitem[Stoddard et~al\mbox{.}(2021)]%
        {Stoddard:2021aa}
\bibfield{author}{\bibinfo{person}{Isak Stoddard}, \bibinfo{person}{Kevin Anderson}, \bibinfo{person}{Stuart Capstick}, \bibinfo{person}{Wim Carton}, \bibinfo{person}{Joanna Depledge}, \bibinfo{person}{Keri Facer}, \bibinfo{person}{Clair Gough}, \bibinfo{person}{Frederic Hache}, \bibinfo{person}{Claire Hoolohan}, \bibinfo{person}{Martin Hultman}, \bibinfo{person}{Niclas H{\"a}llstr{\"o}m}, \bibinfo{person}{Sivan Kartha}, \bibinfo{person}{Sonja Klinsky}, \bibinfo{person}{Magdalena Kuchler}, \bibinfo{person}{Eva L{\"o}vbrand}, \bibinfo{person}{Naghmeh Nasiritousi}, \bibinfo{person}{Peter Newell}, \bibinfo{person}{Glen~P. Peters}, \bibinfo{person}{Youba Sokona}, \bibinfo{person}{Andy Stirling}, \bibinfo{person}{Matthew Stilwell}, \bibinfo{person}{Clive~L. Spash}, {and} \bibinfo{person}{Mariama Williams}.} \bibinfo{year}{2021}\natexlab{}.
\newblock \showarticletitle{Three Decades of Climate Mitigation: Why Haven't We Bent the Global Emissions Curve?}
\newblock \bibinfo{journal}{\emph{Annual Review of Environment and Resources}} \bibinfo{volume}{46}, \bibinfo{number}{1} (\bibinfo{year}{2021}), \bibinfo{pages}{653--689}.
\newblock


\bibitem[Ulrich and Reynolds(2010)]%
        {Ulrich2010}
\bibfield{author}{\bibinfo{person}{Werner Ulrich} {and} \bibinfo{person}{Martin Reynolds}.} \bibinfo{year}{2010}\natexlab{}.
\newblock \showarticletitle{Critical Systems Heuristics}.
\newblock In \bibinfo{booktitle}{\emph{Systems Approaches to Managing Change: A Practical Guide}}, \bibfield{editor}{\bibinfo{person}{Martin Reynolds} {and} \bibinfo{person}{Sue Holwell}} (Eds.). \bibinfo{publisher}{Springer London}, \bibinfo{address}{London}, \bibinfo{pages}{243--292}.
\newblock


\bibitem[{United Nations Educational, Scientific, and Cultural Organization}(2017)]%
        {UNESCO:2017aa}
\bibfield{author}{\bibinfo{person}{{United Nations Educational, Scientific, and Cultural Organization}}.} \bibinfo{year}{2017}\natexlab{}.
\newblock \bibinfo{title}{Education for Sustainable Development Goals -- Learning Objectives}.
\newblock


\bibitem[Venters et~al\mbox{.}(2018)]%
        {VENTERS2018}
\bibfield{author}{\bibinfo{person}{Colin~C. Venters} {et~al\mbox{.}}} \bibinfo{year}{2018}\natexlab{}.
\newblock \showarticletitle{Software sustainability: Research and practice from a software architecture viewpoint}.
\newblock \bibinfo{journal}{\emph{Journal of Systems and Software}}  \bibinfo{volume}{138} (\bibinfo{year}{2018}), \bibinfo{pages}{174--188}.
\newblock
\showISSN{0164-1212}
\urldef\tempurl%
\url{https://doi.org/10.1016/j.jss.2017.12.026}
\showDOI{\tempurl}


\bibitem[Venters et~al\mbox{.}(2023)]%
        {VENTERS2023}
\bibfield{author}{\bibinfo{person}{Colin~C. Venters} {et~al\mbox{.}}} \bibinfo{year}{2023}\natexlab{}.
\newblock \showarticletitle{Sustainable software engineering: Reflections on advances in research and practice}.
\newblock \bibinfo{journal}{\emph{Information and Software Technology}}  \bibinfo{volume}{164} (\bibinfo{year}{2023}), \bibinfo{pages}{107316}.
\newblock
\showISSN{0950-5849}
\urldef\tempurl%
\url{https://doi.org/10.1016/j.infsof.2023.107316}
\showDOI{\tempurl}


\bibitem[Whittle(2019)]%
        {8693084}
\bibfield{author}{\bibinfo{person}{Jon Whittle}.} \bibinfo{year}{2019}\natexlab{}.
\newblock \showarticletitle{Is Your Software Valueless?}
\newblock \bibinfo{journal}{\emph{IEEE Software}} \bibinfo{volume}{36}, \bibinfo{number}{3} (\bibinfo{year}{2019}), \bibinfo{pages}{112--115}.
\newblock


\bibitem[Wiek et~al\mbox{.}(2015)]%
        {Wiek:2015aa}
\bibfield{author}{\bibinfo{person}{Arnim Wiek}, \bibinfo{person}{Michael~J Bernstein}, \bibinfo{person}{Rider~W Foley}, \bibinfo{person}{Matthew Cohen}, \bibinfo{person}{Nigel Forrest}, \bibinfo{person}{Christopher Kuzdas}, \bibinfo{person}{Braden Kay}, {and} \bibinfo{person}{Lauren~Withycombe Keeler}.} \bibinfo{year}{2015}\natexlab{}.
\newblock \showarticletitle{Operationalising competencies in higher education for sustainable development}.
\newblock In \bibinfo{booktitle}{\emph{Routledge handbook of higher education for sustainable development}}. \bibinfo{publisher}{Routledge}, \bibinfo{pages}{265--284}.
\newblock


\end{thebibliography}

\end{document}